\documentclass{ws-ijmpd}
\usepackage[super,compress]{cite}
\begin{document}

\markboth{Piyali Bhar et al.}
{A new class of relativistic model of compact stars of embedding class I}

%
\catchline{}{}{}{}{}
%

\title{A new class of relativistic model of compact stars of embedding class I}

\author{Piyali Bhar}

\address{Department of Mathematics, Government General Degree College, Singur,\\
Hooghly, West Bengal-712409, India\\
piyalibhar90@gmail.com}

\author{Ksh. Newton Singh}
\address{Department of Physics, National Defence Academy, Khadakwasla, Pune-411023, India.\\
ntnphy@gmail.com}

\author{Tuhina Manna}
\address{Department of Commerce (Evening), St. Xavier's College, 30
Mother Teresa Sarani, Kolkata 700016, West Bengal, India\\
tuhinamanna03@gmail.com}

\maketitle

\begin{history}
\received{Day Month Year}
\revised{Day Month Year}
\end{history}

\begin{abstract}\noindent
In the present paper we have constructed a new relativistic anisotropic compact star model having a spherically symmetric metric of embedding class one. Here we have assumed an arbitrary form of metric function $e^{\lambda}$ and solved the Einstein's relativistic field equations with the help of Karmarkar condition for an anisotropic matter distribution. The physical properties of our model such as pressure, density, mass function, surface red-shift, gravitational red-shift are investigated and the stability of the stellar configuration is discussed in details. Our model is free from central singularities and satisfies all energy conditions. The model we present here satisfy the static stability criterion i.e. $dM/d\rho_c>0$ for $0\le \rho_c \le 4.04\times 10^{17}~g/cm^3$ (stable region) and for $\rho_c \ge 4.04\times 10^{17}~g/cm^3$, the region is unstable i.e., $dM/d\rho_c \le 0$.\\

PACS Number(s): 04.40.Nr, 04.20.Jb, 04.20.Dw
\end{abstract}
\maketitle

\section{Introduction}

The physics of compact stars have always drawn the interest to the researchers. The theoretical idea of black holes as exotic objects whose gravitational fields are so strong that even light cannot escape have originated as early as in the $18$th century. But round that time it remained only as a mere textbook curiosity. Compact objects as real astrophysical objects where only considered in the early $1930$'s. Chadwick's discovery of a neutron in 1932 sparked renewed interest in the gravitational collapse of compact stars, following which Baade and  Zwicky suggested the notion of a neutron star while investigating supernovae observations. Most importantly Einstein's theory of general relativity laid the foundation of our understanding of compact stars. Early research works on the structure of compact objects have been done by Oppenheimer, Volkoff and Tolman in $1939$ \cite{Oppenheimer,Tolman}. They successfully derived the equations of relativistic stellar structures from Einstein's field equations and discovered the existence of a limit to the mass of stable degenerate relativistic stars.\par

 In the past few years cooperative efforts of radio and optical astronomers have revealed a great deal of observational data which in turn has provided  physicists new theoretical insights into the physics of compact objects. Fascinating new observational data of the stellar objects Her X-1, Cyg X-2, 4U 1820-30, SAX J 1808.4-3658, 4U 1728-34, PSR 0943+10 and RX J185635-3754 have further contributed to this area of research. More sophisticated data have challenged us to investigate more precise and complicated approach to stellar modeling. In comparison to normal stars of comparable mass, compact objects have much smaller radii and hence, much stronger surface gravity. The interior structures of compact objects have extremely high densities and involve phases which are not yet clearly understood. Such high densities of the order of $10^{15}g~cm^{-3}$ or even higher, as suggested by some recent developments, causes the nuclear fluids in the interior of a compact star to be anisotropic as suggested by Ruderman \cite{Ruderman}. For an anisotropic fluid, the radial and tangential pressures are not equal (i.e. $p_r \neq  p_t$) and the anisotropy factor $\Delta = p_t - p_r$ increases rapidly with the increase in radial distance but vanishes at the centre of our stellar model. The cause of anisotropy may be due to a variety of reasons like  presence of a solid core or type $3A$ fluid or type $P$ superfluid \cite{Kippenhahn} or may result from different kind of phase transition, rotation, magnetic stress, pion condensation etc. \cite{Sokolov,Sawyer}. In an early work Ponce de Le\'{o}n \cite{Ponce} obtained two new exact analytical solutions to Einstein's field equations for static fluid sphere with anisotropic pressures. In another notable work  Herrera \& Santos \cite{Herrera} provided an exhaustive review on the subject of anisotropic fluids. B\"{o}hmer and Harko \cite{Bohmer} obtained upper and lower bounds of the physical parameters like mass-radius ratio, anisotropy, red-shift and total energy for anisotropic fluid in the presence of the cosmological constant. In addition to that, we would like to mention the works of Mak and Harko \cite{Mak} and Sharma and Maharaj \cite{Sharma} which suggest that anisotropy is a sufficient condition in the study of dense nuclear matter. By assuming pressure anisotropy Bhar \cite{pb1,pb2,pb3} obtained anisotropic model of compact star in $(3+1)D$ space-time. Bhar {\em et al.} \cite{pb4} studied the behavior of static spherically symmetric relativistic objects with locally anisotropic matter distribution considering the Tolman VII form for the gravitational potential $g_{rr}$ in curvature coordinates together with the linear relation between the energy density and the radial pressure. A new class of interior solutions for anisotropic stars admitting conformal motion in higher dimensional non-commutative space-time by choosing a particular density distribution function of Lorentzian type as provided by Nazari and Mehdipour \cite{meh} was obtained by Bhar et al \cite{pb5}. A large number of works on the modeling of compact object in the platform of KB metric were done by several researchers \cite{varela,kalam,hos,rah}.\par
In this paper we have investigated a new relativistic model of compact star of embedding class one. Here we have used the well known fact that the {\it n} dimensional manifold {\it $V_n$} can be embedded in a pseudo-Euclidean space of {\it m} dimensions where $ m=n(n+1)/2$. The embedding class of {$V_n$} is the minimum number $m-n$ of extra dimensions which needs to be added to manifold {\it $V_n$}. Note that in a metric of embedding class one the metric functions $\lambda$ and $\nu$ are dependent on each other. Due to such exceptional relationship, it is possible to generate either Schwarzschild interior solution \cite{sch1} or Kohlar-Chao solution \cite{koh1} for a neutral perfect fluid distribution. However, a very recently  Bhar et al. \cite{Piyali1} and Maurya et al. \cite{Maurya1,Maurya2,Maurya3} have investigated an anisotropic compact star model of embedding class one. Some other notable works in this field have been done extensively by Singh et al.\cite{Singh1,Singh2,Singh3,sing1,sing2,sing3}.\par
The paper has been divided in the following sections; in section $2$ the field equations have been solved using the Karmakar condition.In section $3$ a particular model have been discussed, next in section $4$ we have worked on various physical acceptability conditions has must satisfied by our model. The exterior space-time and boundary conditions have been investigated in section $5$. After that in section $6$ the different properties of the solution have been discussed in details. Next in section $7$ we have analyzed the stability of our stellar model in various conditions. Lastly in section $8$ we have discussed some concluding remarks.

\section{Field Equations and Karmarkar's Condition}
In the canonical coordinate $(x^{\mu})\equiv (t,~r,~\theta,~\phi)$, the interior of a static and spherically symmetry object is described by line element
\begin{equation}\label{line}
ds^{2}=e^{\nu(r)}dt^{2}-e^{\lambda(r)}dr^{2}-r^{2}\left(d\theta^{2}+\sin^{2}\theta d\phi^{2} \right)
\end{equation}
Where $\nu$ and $\lambda$ are functions of the radial coordinate `$r$' only.\par
The Einstein field equations for anisotropic fluid distribution are given as (in the unit $G=c=1$)
\begin{equation}
-8\pi T^\mu_\xi = \mathcal{R}^\mu_\xi-{1\over 2}\mathcal{R}~g^\mu_\xi
\end{equation}
where
\begin{eqnarray}
T^\mu_\xi & = & \rho v^\mu v_\xi + p_r \chi_\xi \chi^\mu + p_t(v^\mu v_\xi -\chi_\xi \chi^\mu-g^\mu_\xi) ~,\label{t}
\end{eqnarray}
the energy-momentum tensor, $\mathcal{R}^\mu_\xi$  is the Ricci tensor, $\mathcal{R}$ is the scalar curvature, $p_r$ and $p_t$ denote radial and transverse pressures respectively, $\rho$ the density distribution , $v^\mu$   the four velocity and $\chi^\mu$  is the unit space-like vector in the radial direction.\par
Now for the line element (\ref{line}) and the matter distribution (\ref{t}) in Einstein Field equations (assuming $G=c=1$) take this form,
\begin{eqnarray}
	\frac{1-e^{-\lambda}}{r^{2}}+\frac{e^{-\lambda}\lambda'}{r}&=&8\pi\rho,\label{f1}\\
	\frac{e^{-\lambda}-1}{r^{2}}+\frac{e^{-\lambda}\nu'}{r}&=&8\pi p_{r},\label{f2}\\
	e^{-\lambda}\left(\frac{\nu''}{2}+\frac{\nu'^{2}}{4}-\frac{\nu'\lambda'}{4}+\frac{\nu'-\lambda'}{2r} \right)&=&8\pi p_t.\label{f3}
\end{eqnarray}
where $`\prime'$ represents differentiation with respect to the radial coordinate $r$. Using the Eqs. (\ref{f2}) and (\ref{f3}) we obtain the anisotropy parameter
\begin{eqnarray}
\Delta&=&p_t-p_r \nonumber\\
&&= \frac{e^{-\lambda}}{8\pi}\left[{\nu'' \over 2}-{\lambda' \nu' \over 4}+{\nu'^2 \over 4}-{\nu'+\lambda' \over 2r}+{e^\lambda-1 \over r^2}\right]\nonumber\\\label{del}
\end{eqnarray}
Now if the space-time (\ref{line}) satisfies the Karmarkar condition \cite{kar}
\begin{equation}
\mathcal{R}_{1414}=\frac{\mathcal{R}_{1212}\mathcal{R}_{3434}+ \mathcal{R}_{1224}\mathcal{R}_{1334}}{\mathcal{R}_{2323}}\label{con}
\end{equation}
with $\mathcal{R}_{2323}\neq 0$ \cite{pandey}, it represents the space time of embedding class $1$.\\
Now the above components of $\mathcal{R}_{hijk}$ for metric (\ref{line}) are:
\begin{eqnarray}
\mathcal{R}_{2323}&=&\,r^2\,\sin^{2}\theta~\left[1-e^{-\lambda}\right] \nonumber\\
\mathcal{R}_{1212}&=&\frac{1}{2}\,\lambda'\,r \nonumber\\
\mathcal{R}_{1334} &=& R_{1224} \sin^2 \theta = 0   \nonumber\\
\mathcal{R}_{1414}&=& -e^{\nu}\left[\frac{1}{2}~\nu''+\frac{1}{4}~{\nu'}^{2}-\frac{1}{4}~\lambda'~\nu'\right] \nonumber\\
\mathcal{R}_{3434}&=& -\frac{r}{2}\sin^{2}\theta~ \nu'e^{\nu-\lambda} \nonumber
\end{eqnarray}
Using the above expression eq.(\ref{con}) gives the following differential equation
\begin{equation}
{2\nu'' \over \nu'}+\nu'={\lambda' e^\lambda \over e^\lambda-1}\label{dif1}
\end{equation}
On integration we get the following relationship between the metric potentials $\nu$ and $\lambda$ as
\begin{equation}
e^{\nu}=\left(A+B\int \sqrt{e^{\lambda}-1}~dr\right)^2\label{nu1}
\end{equation}
where $A$ and $B$ are constants of integration.
By using (\ref{nu1}) we can rewrite (\ref{del}) as
\begin{eqnarray}
\Delta = {\nu' \over 4e^\lambda}\left[{2\over r}-{\lambda' \over e^\lambda-1}\right]~\left[{\nu' e^\nu \over 2rB^2}-1\right] \label{del1}
\end{eqnarray}

We have to solve the Einstein field equations (\ref{f1})-(\ref{f3}) with the help of equation (\ref{nu1}). One can notice that we have four equations with $5$ unknowns namely $\lambda,\nu,\rho,p_r$ and $p_t$.
\begin{figure}[!htb]\centering
       \begin {minipage}{0.6\textwidth}
    \includegraphics[width=\linewidth]{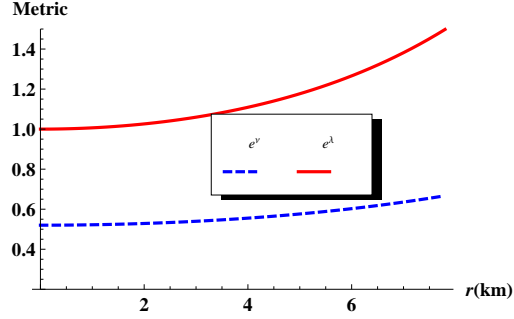}
     \end{minipage}
    \caption{The metric potentials are plotted against $r$ by taking $a=0.0.08,\,b=-0.00098,\,A=-0.786839,\,B=0.0369469$ for the compact star 4U 1538-52}\label{metric}
    \end{figure}
\begin{figure}[!htb]\centering
       \begin {minipage}{0.6\textwidth}
    \includegraphics[width=\linewidth]{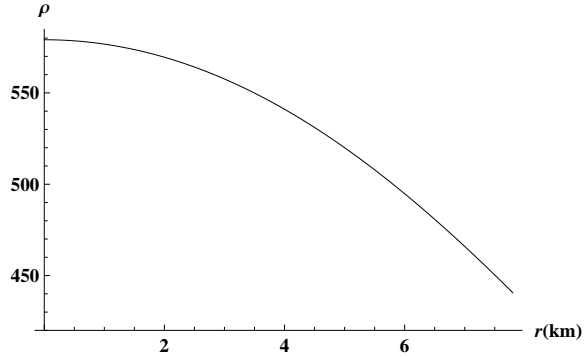}
     \end{minipage}
    \caption{Matter density is plotted against $r$ for the values mentioned in Fig. \ref{metric}.}\label{rho1}
\end{figure}

\section{A particular Model}
To generate the model let us assume the expression for the $g_{rr}$ metric
potential as,
\begin{equation}\label{elambda}
e^{\lambda}=1 + \frac{a^2 r^2}{(1 + b r^2)^4}
\end{equation}
Where $a$ and $b$ are constants having the dimensions {\em length$^{-1}$} and {\em length$^{-2}$} respectively. In a very recent work Singh \& Pant \cite{sing1} obtained a model of compact star of embedding class I by choosing the metric potential as $e^{\lambda}=1 + a^2 r^2(1 + b r^2)^n$. They have only analyze the results for positive values of $n$. In our present paper we follow the above {\em ansatz} with $n=-4$.\\
Solving eqns. (\ref{nu1}) and (\ref{elambda}) we obtain the expression for the metric coefficient $e^{\nu}$ as,
\begin{equation}\label{enu}
e^{\nu}=\left[A -\frac{a B}{2 b (1 + b r^2)}\right]^2
\end{equation}
Using eqns.(\ref{elambda}) and (\ref{enu})the expressions for matter density, radial and transverse pressure are obtained as,
\begin{eqnarray}
8\pi \rho&=&\frac{a^2\big[a^2 r^2 +(1 + b r^2)^3 (3-5 b r^2)\big]}{\big[a^2 r^2 + (1 +
    b r^2)^4\big]^2} \label{rho}\\
8\pi p_r&=&\frac{a \big[a^2 B - 2 a A b (1 + b r^2) + 4 b B (1 + b r^2)^3\big]}{\big[
   2 A b (1 + b r^2)-aB\big]\big[a^2 r^2 + (1 + b r^2)^4\big]} \label{pr}\\
8\pi p_t&=&\frac{a (1 + b r^2)^3}{\big[
2 A b (1 + b r^2)-a B\big]\big[a^2 r^2 + (1 + b r^2)^4\big]^2}\times\nonumber\\
&&\bigg[a^2 B (1 - b r^2) +
   4 b B f_1(r)-
   2 a A b f_2(r)\bigg]\label{pt}
\end{eqnarray}
where,
\begin{eqnarray*}
f_1(r)&=& (1 - b r^2) (1 + b r^2)^3\\
f_2(r)&=&(1 + b r^2) (1 - 3 b r^2)
\end{eqnarray*}

\begin{figure}[!htb]\centering
       \begin {minipage}{0.6\textwidth}
    \includegraphics[width=\linewidth]{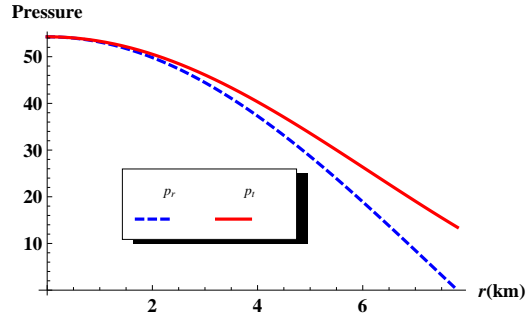}
     \end{minipage}
    \caption{The radial and transverse pressure are plotted against $r$ for the values mentioned in Fig. \ref{metric}.}\label{pr1}
\end{figure}

\begin{figure}[!htb]\centering
       \begin {minipage}{0.6\textwidth}
    \includegraphics[width=\linewidth]{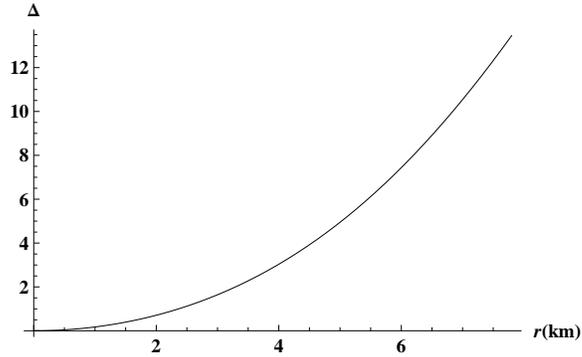}
     \end{minipage}
    \caption{The anisotropic factor is plotted against $r$ for the values mentioned in Fig. \ref{metric}.}\label{delta}
\end{figure}

\begin{figure}[!htb]\centering
       \begin {minipage}{0.6\textwidth}
    \includegraphics[width=\linewidth]{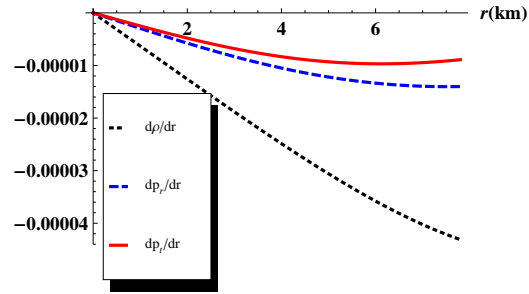}
     \end{minipage}
    \caption{The density and pressure gradients are plotted against $r$ for the values mentioned in Fig. \ref{metric}.}\label{dp}
\end{figure}

\begin{figure}[!htb]\centering
       \begin {minipage}{0.6\textwidth}
    \includegraphics[width=\linewidth]{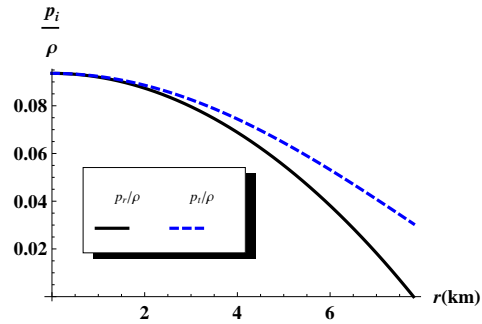}
     \end{minipage}
    \caption{$p_r/\rho$ and $p_t/\rho$ are plotted against $r$ for the values mentioned in Fig. \ref{metric}.}\label{ratio}
\end{figure}

\begin{figure}[!htb]\centering
       \begin {minipage}{0.6\textwidth}
    \includegraphics[width=\linewidth]{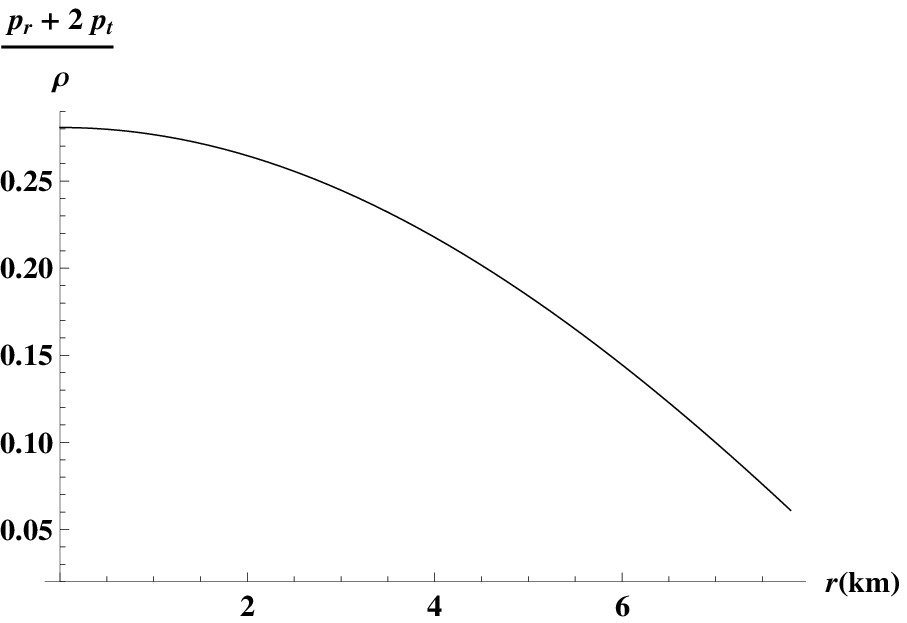}
     \end{minipage}
    \caption{$(p_r+2p_t)/\rho$ is plotted against $r$ for the values mentioned in Fig. \ref{metric}.}\label{trace}
\end{figure}

\section{Physical acceptability conditions}\label{Sec4}

For the well behaved nature of the solution, the following conditions should be satisfied \cite{abreu}:
\begin{description}
  \item[(i)] The metric potentials should be free from singularities inside the radius of the star moreover the fluid sphere should satisfy $e^{\nu(0)}=$ constant, and $e^{-\lambda(0)}=1$.
  \item[(ii)] The density $\rho$ and pressures $p_r,\,p_t$  should be positive inside the fluid configuration and should decreasing outward.
  \item[(iii)] The radial pressure $p_r$ must be vanishing but the tangential pressure $p_t$  may not necessarily vanish at the boundary $r=r_\Sigma$. However, the radial pressure is equal to the tangential pressure at the center of the fluid sphere, i.e., pressure anisotropy vanishes at the center, $\Delta(0)=0$ \cite{bl,iva} and $\displaystyle\Delta(r=r_\Sigma)=p_t(r_\Sigma)>0$ \cite{bh}.
  \item[(iv)] The radial pressure gradient $dp_r/dr\leq0$~for $0\leq r \leq r_\Sigma$.
  \item[(v)] The density gradient $d\rho/dr\leq0$ for $0\leq r\leq r_\Sigma$.
  \item[(vi)] A physically acceptable fluid sphere must satisfy the causality conditions, the radial and tangential adiabatic speeds of sound should less than the speed of light. In the unit $c=1$ the causality conditions take the form $0<v_{sr}^{2}=dp_r/d\rho\leq 1$ and $0<v_{st}^{2}=dp_t/d\rho\leq 1$.
  \item[(vii)] The interior solution should satisfy either
              \begin{itemize}
                \item strong energy condition (SEC) $\rho-p_r-2p_t\geq0,\,\rho-p_r\geq0,\,\rho-p_t\geq0$ or
                \item dominant energy condition (DEC) $\rho\geq p_r$ and $\rho\geq p_t$.
              \end{itemize}
  \item[(viii)] The interior solution should continuously match with the exterior Schwarzschild solution.
\end{description}
Conditions (iv) and (v) imply that pressure and density should be maximum at the center and monotonically decreasing towards the surface.\par

\begin{figure}[!htb]\centering
       \begin {minipage}{0.6\textwidth}
    \includegraphics[width=\linewidth]{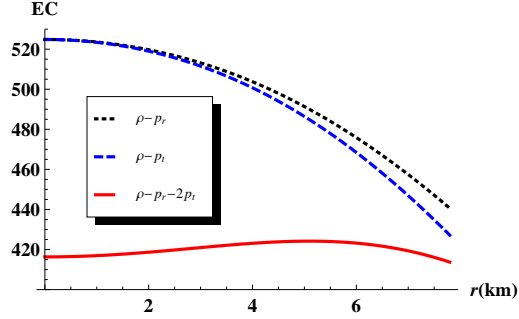}
     \end{minipage}
    \caption{The energy conditions are potted against $r$ for the values mentioned in Fig. \ref{metric}.}\label{ec}
\end{figure}

\begin{figure}[!htb]\centering
       \begin {minipage}{0.6\textwidth}
    \includegraphics[width=\linewidth]{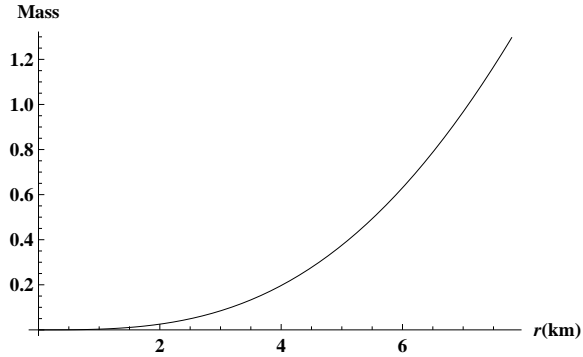}
     \end{minipage}
    \caption{The mass function is plotted against $r$ for the values mentioned in Fig. \ref{metric}.}\label{mass1}
\end{figure}

\section{Exterior space-time and boundary condition}
To fix the values of the constants $a,\,b,\,A$ and $B$ we match our interior spacetime to the exterior Schwarzschild line element given by,
\begin{eqnarray}
ds^{2}&=&\left(1-\frac{2m}{r}\right)dt^{2}-\left(1-\frac{2m}{r}\right)^{-1}dr^{2}\nonumber\\
&&-r^{2}(d\theta^{2}+\sin^{2}\theta d\phi^{2})
\end{eqnarray}
outside the event horizon $r>2m$, $m$ being the mass of the black hole.\par
Using the continuity of the metric coefficients $e^{\nu},e^{\lambda}$ across the boundary we get the following three equations
\begin{eqnarray}
1 + \frac{a^2 r_{\Sigma}^2}{(1 + b r_{\Sigma}^2)^4}&=&\left(1-\frac{2m}{r_{\Sigma}}\right)^{-1} \label{b1}\\
1-\frac{2m}{r_{\Sigma}}&=&\left[A -\frac{a B}{2 b (1 + b r_{\Sigma}^2)}\right]^2 \label{b2}
\end{eqnarray}
and $p_r(r=r_{\Sigma})=0$ gives,
\begin{equation}\label{b3}
a^2 B - 2 a A b (1 + b r_{\Sigma}^2) + 4 b B (1 + b r_{\Sigma}^2)^3=0
\end{equation}
Solving eqns. (\ref{b1})-(\ref{b3})we get,
\begin{eqnarray}
a&=& \frac{(1+br_{\Sigma}^2)^2}{r_{\Sigma}}\sqrt{\frac{2m/r_{\Sigma}}{1-2m/r_{\Sigma}}} \\
B^2&=& \frac{m}{2r_{\Sigma}^3} \\
A&=& \frac{\big[a^2+4b(1+br_{\Sigma}^2)^3\big] B}{2ab(1+br_{\Sigma}^2)}
\end{eqnarray}
\begin{figure}[!htb]\centering
       \begin {minipage}{0.6\textwidth}
    \includegraphics[width=\linewidth]{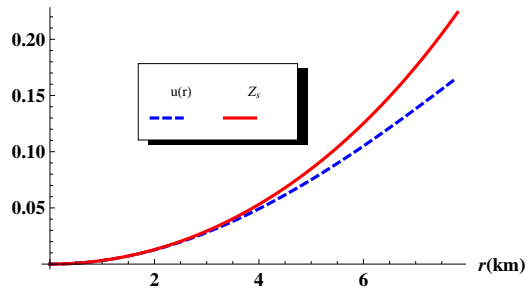}
     \end{minipage}
    \caption{The compactness factor and surface redshift are plotted against $r$ for the values mentioned in Fig. \ref{metric}.}\label{zs}
\end{figure}

\begin{figure}[!htb]\centering
       \begin {minipage}{0.6\textwidth}
    \includegraphics[width=\linewidth]{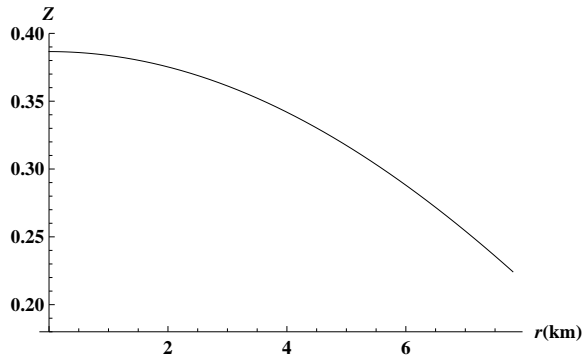}
     \end{minipage}
    \caption{The gravitational redshift factor is plotted against $r$ for the values mentioned in Fig. \ref{metric}.}\label{z}
\end{figure}

\section{Properties of the solution}
\subsection{Singularity free nature of the model parameter}
At the center of the star the expressions for metric potentials are obtained as,
\[e^{\lambda}|_{r=0}=1,\, e^{\nu}|_{r=0}=\left(A-\frac{aB}{2b}\right)^2\]
which are constants and
\[(e^{\lambda})'|_{r=0}=0,\, (e^{\nu})'|_{r=0}=0.\]
The central density and central pressure are obtained as,
\begin{eqnarray}
8\pi \rho_c&=&3a^2\\
\label{pc}8\pi p_r(r=0)=8\pi p_t(r=0)&=&\frac{a \big[a^2 B - 2 a A b + 4 b B \big]}{\big[
   2 A b -aB\big]}\nonumber\\
\end{eqnarray}
To satisfy Zeldovich's condition at the interior, $p_r/\rho$ at
the center must be $\leq 1$. Therefore
\begin{equation}\label{zel}
\frac{a^2 B - 2 a A b + 4 b B}{3a\big[
   2 A b -aB\big]}\leq1
\end{equation}
From eqns. (\ref{pc}) and (\ref{zel}) we get,
\begin{eqnarray}
  \frac{a^2+b}{2ab} \leq \frac{A}{B} < \frac{a^2+4b}{2ab}
\end{eqnarray}

Differentiating eqns. (\ref{rho})-(\ref{pt}) we get the density and pressure gradient as,
\begin{eqnarray}
8\pi\frac{d\rho}{dr}&=&-\frac{2 a^2 r f_3(r)}{(a^2 r^2 + (1 + b r^2)^4)^3}\label{drho}\\
8\pi\frac{dp_r}{dr}&=&\frac{2ar}{[a^2 r^2+(1 + b r^2)^4]^2 \{a B -
   2 A b (1 + b r^2)\}^2}\times   \nonumber\\
&&\bigg[a^5 B^2 - 4 a^4 A b B (1 + b r^2) - 16 A b^3 B (1 + b r^2)^7\nonumber\\
 &&- 8 a^2 A b^2B f_4(r)
   +4 a^3 b (1 + b r^2)^2 f_5(r) +\nonumber\\
   &&4 a b^2 (1 + b r^2)^5 f_6(r) \bigg]\label{dpr}\\
8\pi\frac{dp_t}{dr}&=&\frac{4 a r (1 +
   b r^2)^2}{\big[a B -
   2 A b (1 + b r^2)\big]^2 \big[a^2 r^2 + (1 + b r^2)^4\big]^3}\times\nonumber\\
   &&\bigg[-8 A b^3 B (2-b r^2) (1 + b r^2)^8 -
  2 a^2 A b^2 B f_7(r) \nonumber \\
  && + a^5 B^2 \big[1 - b r^2 (1-b r^2)\big] -
  2 a b^2 (1 + b r^2)^6 f_8(r) \nonumber \\
  &&-2 a^4 A b B (1 + b r^2)\big[2-b r^2(3-5 b r^2)\big] \nonumber\\
  &&+a^3 b (1 + b r^2)^2 f_9(r)\bigg]\label{dpt}
\end{eqnarray}

where,
\begin{eqnarray*}
f_3(r)&=&a^4 r^2+
  20 b (1-b r^2) (1 + b r^2)^6 +\\
   &&(a + a b r^2)^2[5-
     b r^2 (2-17 b r^2)]\\
f_4(r)&=&(1 + b r^2)^3 (3 + b r^2)\\
f_5(r)&=& A^2 b + B^2 (2 - b r^2)\\
f_6(r)&=&4 A^2 b + B^2 (1 + b r^2)\\
f_7(r)&=&(1 + b r^2)^4 \big[11 - b r^2(8-b r^2)\big]\\
f_8(r)&=&B^2 (-3 + b r^2) (1 + b r^2) +
     4 A^2 b (-2 + 3 b r^2)\\
f_9(r)&=&4 A^2 b (1- b r^2(2-3 b r^2))\\
     &&+B^2(1 + b r^2)\big[7-b r^2(9-8 b r^2)\big]
\end{eqnarray*}

The profiles of the metric coefficients, matter density, radial and transverse pressure are plotted in Figs. \ref{metric}, \ref{rho1} and fig. \ref{pr1} respectively which shows that metric coefficients, $\rho$, $p_r$ and $p_t$ all are positive and free from central singularity. The radial pressure $p_r$ vanishes at the boundary of the star whereas the transverse pressure and matter density are still positive inside the stellar interior as well as at the boundary. $\rho,\,p_r$ and $p_t$ all are monotonic decreasing function of $r$. The monotonic decreasing condition is verified by Fig.~\ref{dp}. The anisotropic factor 	 is plotted against $r$ in Fig.~\ref{delta}. The profile shows that $\Delta>0$ inside the stellar configuration and therefore the anisotropic force is repulsive in nature and according to Gokhroo and Mehra \cite{gm} it is necessary to construct the compact object. Both $p_r/\rho$ and $p_t/\rho$ are monotonic decreasing function of r and lies in the range $0<p_r/\rho,\,p_t/\rho<1$ (Fig.\ref{ratio}) which verifies that the underlying fluid distribution is non-exotic in nature. $(p_r+2p_t)/\rho$ is plotted against $r$ in fig. \ref{trace}, which is monotonic decreasing function of $r$ and $0<(p_r+2p_t)/\rho<1.$ All the energy conditions are satisfied by our present model which is shown in fig.~\ref{ec}.

\subsection{Mass-radius relation}
Using the relationship $e^{-\lambda}=1-\frac{2m}{r}$ the mass function is obtained as
\begin{equation}\label{mass}
m(r)=\frac{a^2r^3}{2\big[(1+br^2)^4+a^2r^2\big]},
\end{equation}
the compactness factor and surface red-shift are obtained as,
\begin{eqnarray}
  u(r) &=& \frac{a^2r^2}{2\big[(1+br^2)^4+a^2r^2\big]} \\
  z_s &=& \Big[\frac{(1+br_\Sigma^2)^4}{(1+br_\Sigma^2)^4+a^2r_\Sigma^2}\Big]^{-\frac{1}{2}}-1
\end{eqnarray}

The gravitational red-shift of the stellar configuration is given by
\begin{eqnarray}
z &=& e^{-\nu/2}-1 = \left[A - \frac{a B}{2 b (1 + b r^2)}\right]^{-1}-1
\end{eqnarray}
The profile of mass function is plotted in fig.~\ref{mass1}, which is monotonic increasing in nature and regular at the center of the star as well as positive inside the stellar interior. Following the concept of Buchdahl \cite{buch} for a compact star the ratio of mass to the radius can not be arbitrarily large. The maximally allowable mass-to-radius ratio $(M/r_{\Sigma})$ for an isotropic fluid sphere should lie in the range $ 2M/r_{\Sigma} <8/9$ (in the units $c=G=1$). The compactification factor and surface red-shift are plotted in figs. \ref{zs}. Both the profiles are monotonic increasing in nature. On the other hand the profile of gravitational red-shift is plotted against $r$ in fig.~\ref{z}, which is monotonically decreasing in nature.

\section{Stability}

\subsection{Stability under three different forces}
For checking the static equilibrium of our present model under three forces $viz$ gravitational force, hydrostatics force and anisotropic force we consider the following equation
\begin{equation}\label{tov1}
-\frac{M_G(r)(\rho+p_r)}{r}e^{\frac{\nu-\lambda}{2}}-\frac{dp_r}{dr}+\frac{2}{r}(p_t-p_r)=0,
\end{equation}
proposed by Tolman-Oppenheimer-Volkoff and called the TOV equation and $M_G(r)$ can derived from the Tolman-Whittaker
formula and the Einstein's field equations and is defined by
\begin{equation}
M_G(r)=\frac{1}{2}re^{\frac{\lambda-\nu}{2}}~\nu'
\end{equation}
represents the gravitational mass within the radius $r$.

Plugging the expression of $M_G(r)$ in equation $(\ref{tov1})$, we obtain
\begin{equation}\label{tov2}
-\frac{\nu'}{2}(\rho+p_r)-\frac{dp_r}{dr}+\frac{2}{r}(p_t-p_r)=0.
\end{equation}
The above equation can be rewritten as,
\begin{equation}
F_g+F_h+F_a=0,
\end{equation}
\begin{figure}[!htb]\centering
       \begin {minipage}{0.6\textwidth}
    \includegraphics[width=\linewidth]{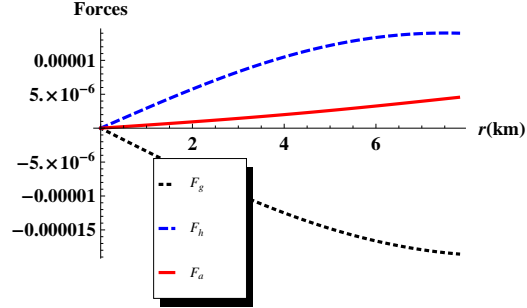}
     \end{minipage}
    \caption{The variation of three different forces acting on the system are potted against $r$ for the values mentioned in Fig. \ref{metric}.}\label{tov}
\end{figure}
where $F_g, F_h$ and $F_a$ represents the gravitational, hydrostatics and anisotropic forces respectively.
Using the Eqs. (\ref{rho})-(\ref{pt}), the expression for $F_g,F_h$ and $F_a$ can be written as,

\begin{eqnarray}
F_g &=&-\frac{\nu'}{2}(\rho+p_r)\nonumber\\
&=& -4 b B r (a + a b r^2)^2 \frac{f_{10}(r)}{f_{11}(r)}\\
F_h &=&-\frac{dp_r}{dr}\\
F_a &=&\frac{2\Delta}{r}
\end{eqnarray}
where,
\begin{eqnarray*}
f_{10}(r)&=&2 b B (1 + b r^2)^4 - 2 a A b (1 + b r^2) (-1 + 3 b r^2) \\
&&+a^2 B (-1 + 5 b r^2)\\
f_{11}(r)&=&[a B - 2 A b (1 + b r^2)]^2 [a^2 r^2 + (1 + b r^2)^4]^2
\end{eqnarray*}

The profile of three different forces are plotted in fig. \ref{tov}. The figure shows that hydrostatics and anisotropic force are positive and is dominated by the gravitational force which is negative to keep the system in static equilibrium.

\subsection{Relativistic adiabatic index}

For a relativistic anisotropic sphere the stability is related to the adiabatic index $\Gamma$, the ratio of two specific heats, defined by Chan et al.\cite{chan1},
\begin{equation}
\Gamma_r=\frac{\rho+p_r}{p_r}\frac{dp_r}{d\rho} ~~;~~ \Gamma_t=\frac{\rho+p_t}{p_t}\frac{dp_t}{d\rho}
\end{equation}

Now $\Gamma>4/3$ gives the condition for the stability of a Newtonian sphere and $\Gamma =4/3$ being the condition for a neutral equilibrium proposed by Bondi \cite{bondi64}. This condition changes for a relativistic isotropic sphere due to the regenerative effect of pressure, which renders the sphere more unstable. For an anisotropic general relativistic sphere the situation becomes more complicated, because the stability will depend on the type of anisotropy.

\begin{figure}[!htb]\centering
       \begin {minipage}{0.6\textwidth}
    \includegraphics[width=\linewidth]{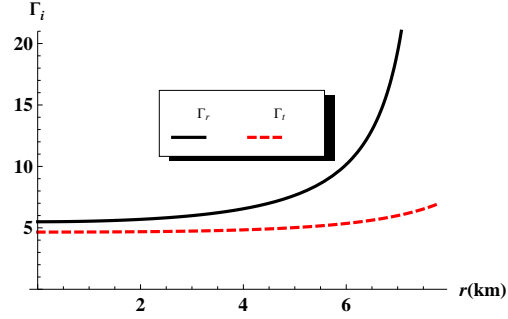}
     \end{minipage}
    \caption{The relativistic adiabatic indices are potted against $r$ for the values mentioned in Fig. \ref{metric}.}\label{f9}
\end{figure}

\subsection{Causality condition}
The radial and tangential speeds of sound for our model of compact star are obtained as,
\begin{eqnarray}
  v_r^2 &=& \frac{dp_r}{d\rho}=\Big(\frac{dp_r}{dr}\Big)/\Big(\frac{d\rho}{dr}\Big) \\
  v_t^2 &=& \frac{dp_t}{d\rho}=\Big(\frac{dp_t}{dr}\Big)/\Big(\frac{d\rho}{dr}\Big)
\end{eqnarray}
\begin{figure}[!htb]\centering
       \begin {minipage}{0.5\textwidth}
    \includegraphics[width=\linewidth]{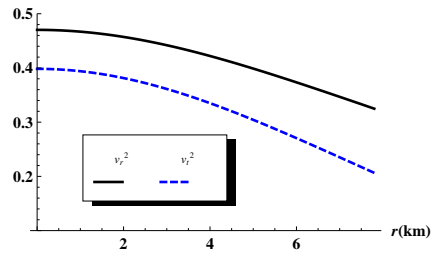}
     \end{minipage}
    \caption{The stability factor are potted against $r$ for the values mentioned in Fig. \ref{metric}.}\label{sv}
\end{figure}

\begin{figure}[!htb]\centering
       \begin {minipage}{0.6\textwidth}
    \includegraphics[width=\linewidth]{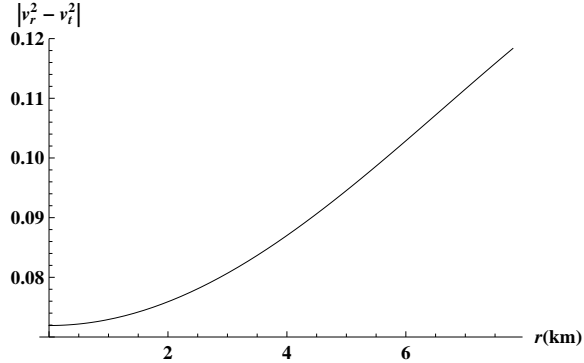}
     \end{minipage}
    \caption{The stability factor are potted against $r$ for the values mentioned in Fig. \ref{metric}.}\label{sv1}
\end{figure}

Where the expression for $\frac{d\rho}{dr},\,\frac{dp_r}{dr}$ and $\frac{dp_t}{dr}$ are shown in the expressions (\ref{drho})-(\ref{dpt}). For a physically acceptable model of relativistic anisotropic star the radial and transverse velocity of sound should lie in the range $0<v_r^2,v_t^2<1$ which is known as causality condition. With the help of graphical representation we have shown that our model satisfies the causality condition (please refer fig. \ref{sv})
To check the stability of anisotropic stars under the radial perturbations Herrera \cite{her} introduced the concept of
``cracking" and using this concept of cracking it was proved by Abreu et al. \cite{abreu} that the region of an anisotropic fluid sphere where $-1 \leq v_t^2-v_r^2\leq0$ is potentially stable but the region where $0 < v_t^2-v_r^2 \leq 1$ is potentially unstable. From Fig.\ref{sv} it is clear that our model satisfies this condition. So we conclude that our model is potentially stable. Moreover $0 <v_r^2 < 1$ and $0 <v_t^2 < 1$ therefore according to Andr\'{e}asson\cite{andre} $|v_r^2 - v_t^2|<1$ which is also clear from Fig.\ref{sv1}.

\begin{figure}[!htb]\centering
       \begin {minipage}{0.6\textwidth}
    \includegraphics[width=\linewidth]{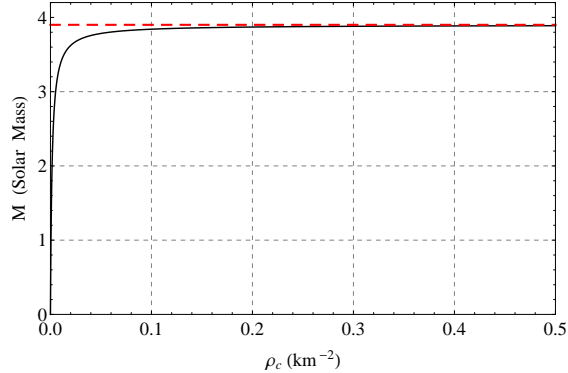}
     \end{minipage}
    \caption{The variation mass with central density (0$-6.74\times 10^{17}~g/cm^3$) for the values mentioned in Fig. \ref{metric} and by assuming $R=7.8~km$. The mass becomes saturated to 3.9 $M_\odot$ when the $\rho_c$ reaches about $4.04\times 10^{17}~g/cm^3$ and onward.}\label{mrho}
\end{figure}

\subsection{Harrison-Zeldovich-Novikov static stability criterion}

Chandrasekhar \cite{chan64}, Harrison \cite{har65} etc. determined the eigen-frequencies of all the fundamental modes to analyzed the stability of stars. However, Harrison \cite{har65} and Zeldovich \& Novikov \cite{zel} simplify into much simpler calculations and reduced it to much more simpler formalism. They have assumed that the adiabatic index of a pulsating star is same as in a slowly deformed matter. This leads to a stable configuration only if the mass of the star is increasing with central density i.e. $dM/d\rho_c > 0$ and unstable if $dM/d\rho_c < 0$ or $dM/d\rho_c = 0$.

In our solution, the mass as a function of central density can be written as
\begin{equation}
M = {8\pi \rho_c R^3/3 \over 2\big(8\pi \rho_c R^2/3+\{1+bR^2\}^4 \big)}\label{mrho1}
\end{equation}
which gives us (for a given radius)

\begin{eqnarray}
{d M \over d \rho_c} &=& 12 \pi  R^3 (b R^2+1)^4 \Big[3 b^4 R^8+12 b^3 R^6+18 b^2 R^4 \nonumber \\
&& +12 b R^2+8 \pi  R^2 \rho_c+3\Big]^{-2}> 0.
\end{eqnarray}

This condition can be further confirmed by Fig. \ref{mrho}. However, when the mass becomes saturated with increase in central density i.e. $dM/d\rho_c=0$ or decrease with increase in $\rho_c$ i.e. $dM/d\rho_c<0$, the unstable configuration is triggered.

\begin{table}[h]
\tbl{The values of the constants calculated from our model for few well-know compact star candidates.}
{\begin{tabular}{@{}cccccc@{}} \toprule
Compact Star &  $a$ &$b$&$A$&$B$ \\
\colrule
4U~1538-52&0.08&-0.00098&-0.786839&0.0369469  \\
PSR~J1614-2230 &0.10&-0.00116&-1.75292&0.0539164 \\
Vela~X-1 &0.09&-0.00125&-1.13613&0.0489534 \\
Cen-X3 &0.09&-0.00088&-1.60872&0.0444594\\
\botrule
\end{tabular} \label{ta1}}
\end{table}

\begin{table}[h]
\tbl{Optimization of masses and radii of few well-know compact star candidates.}
{\begin{tabular}{@{}cccccc@{}} \toprule
Compact Star &$R~$(km)  & $M/M_{\odot}$ &R& $M/M_{\odot}$&References \\
&observed&observed&calculated&calculated\\
\colrule
4U~1538-52&$7.866\pm0.21$&$0.87\pm0.07$&7.8&0.88&Gangopadhyay et al.\cite{gan} \\
PSR~J1614-2230 &$9.69\pm0.2$&$1.97\pm0.04$&9.7& 1.97&Gangopadhyay et al.\cite{gan} \\
Vela~X-1 &$9.56\pm0.08$&$1.77\pm0.08$&9.56 & 1.77&Gangopadhyay et al.\cite{gan}\\
Cen-X3 &$9.178\pm0.13$&$1.49\pm0.08$&9.18&1.5&Gangopadhyay et al.\cite{gan}\\
\botrule
\end{tabular} \label{ta1}}
\end{table}

\begin{table}[h]
\tbl{central density, surface density, central pressure, compactness and surface redshift of few well-know compact star candidates calculated from our model.}
{\begin{tabular}{@{}cccccc@{}} \toprule
Compact Star  & central density   &Surface density& central pressure   & compactness & surface redshift\\
             &        $gm/cm^3$    &$gm/cm^3$&  $dyne/cm^{2}$       &  &   \\
\colrule
4U 1538-52& $1.03\times10^{15}$ &$7.84\times10^{14}$&$8.192\times10^{35}$&0.166 &0.224\\
PSR~ J1614-2230& $1.611\times10^{15}$&$7.502\times10^{14}$&$4.42\times10^{35}$& 0.299&0.579\\
Vela~X-12 & $1.30\times10^{15}$&$7.62\times10^{14}$&$2.88\times10^{35}$& 0.273&0.484\\
Cen X - 3&$1.30\times10^{15}$&$7.26\times10^{14}$&$1.90\times10^{35}$&0.241&0.389\\
\botrule
\end{tabular} \label{ta1}}
\end{table}

\begin{figure}[!htb]\centering
       \begin {minipage}{0.6\textwidth}
    \includegraphics[width=\linewidth]{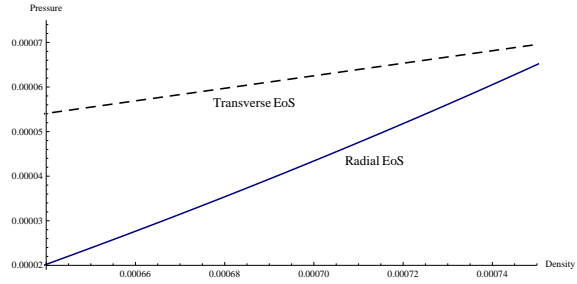}
     \end{minipage}
    \caption{The variation of radial and transverse pressure with respect to density are plotted against $r$. }\label{eos1}
\end{figure}

\section{Discussion}

A perfect fluid distribution satisfying Karmarkar condition is either given by the Schwarzschild interior solution \cite{sch1} or Kohlar-Chao solution \cite{koh1}. To avoid such a perfect fluid distribution we have taken a different form of metric potential $e^\lambda$ along with some anisotropy in pressure and then proceeded with the Karmarkar condition. Some of the key features of the anisotropic stellar model are as follows:
\begin{description}
  \item[(i)] The metric potentials are regular and monotone increasing away from the center of the star (Fig. \ref{metric}). They are free from singularities inside the radius of the star and moreover the fluid sphere satisfy $e^{\nu}(r=0)=$ constant, and $e^{-\lambda}(r=0)=1$ at the stellar center.
  \item[(ii)] From the profiles of  density $\rho$ and pressures $p_r,~p_t$  in Figs.  \ref{rho1} and fig. \ref{pr1} respectively, we have found that they are positive inside the fluid configuration and free from any central singularity.
  \item[(iii)] The radial pressure $p_r$ vanishes at the boundary of the star but the tangential pressure $P_t$  remain positive at the stellar center and at the boundary $r=r_\Sigma$. However, the radial pressure is equal to the tangential pressure at the center of the fluid sphere, i.e., pressure anisotropy factor vanishes at the center, $\Delta(r=0)=0$ as is evident from the profile of anisotropy $\Delta$ in Fig.~\ref{delta}. Also one can note from it that $\Delta>0$ inside the stellar configuration causing the anisotropic force to be repulsive in nature.
  \item[(iv)] The radial pressure gradient $dp_r/dr\leq0$~for $0\leq r \leq r_\Sigma$. The density gradient $d\rho/dr\leq0$ for $0\leq r\leq r_\Sigma$. Both $p_r/\rho$ and $p_t/\rho$ are monotonic decreasing function of r and lies in the range
$0<p_r/\rho,\,p_t/\rho<1$ (fig.~\ref{ratio}) which verifies that the underlying fluid distribution is non-exotic in nature.
  \item[(v)] All the energy conditions are satisfied by our present model which is shown in fig.~\ref{ec}. The ratio of $(p_r + 2p_t)/\rho$ plotted in Fig.~\ref{trace} is monotonic decreasing function of $r$ and less than $1$.
\item[(vi)] The profiles of mass function suggest that it is positive, regular and monotone increasing in nature.The compactification factor and surface red-shift are monotone increasing while the profile of gravitational red-shift suggest that it is monotone decreasing.
\item[(vii)]  The mass of the stellar object increases with the increase in central density (for a given radius) or equivalently $dM/d\rho_c<0$ and thus the static stability criterion hold good till $\rho_c=4.04\times 10^{17}~g/cm^3$. This imply that our solution is not only well-behaved but also represents static and stable. The mass becomes saturated to 3.9 $M_\odot$ when the $\rho_c$ reaches about $4.04\times 10^{17}~g/cm^3$ and onward. Hence the region where $\rho_c>4.04\times 10^{17}~g/cm^3$ satisfy $dM/d\rho=0$ and thus belong to unstable region.
\end{description}
One can note that to solve the Einstein's field equations we have not chosen a particular equation of state. Instead we have assumed $g_{rr}$ and used Karmakar condition to find the another metric coefficient. Then by plugging the expressions of the metric coefficients into the Einstein's field equations we are able to find the expressions for $\rho,\,p_r$ and $p_t$ using an algebraic calculation without solving any differential equations. In the earlier works where the researchers used the KB metric to model a compact stars also found the the expressions for $\rho,\,p_r$ and $p_t$ using an algebraic calculations only \cite{pb1,pb2,varela,kalam,hos,rah}. From the expressions of $\rho,\,p_r$ and $p_t$ we can't obtain a particular equation of state , i.e., a relationship between the radial pressure $p_r$ and the matter density $\rho$ or between the transverse pressure $p_t$ and the matter density $\rho$. To see the trend of the EoS, we plot a graph for $p_r$ and $p_t$ vs $\rho$ in Fig.~\ref{eos1}.\par
Throughout the paper all the plots are drawn for the compact star 4U 1538-52. We have calculated the values of the constants for some well-known compact stars which are presented in table~1. Recently an improved
method was used by Rawls et al.\cite{raw} for determining the mass of neutron stars such as
Vela X-1, SMC X-1, Cen X-3 etc. in eclipsing X-ray pulsar
binaries. To analyze the published
data for these systems they used a numerical code based on Roche geometry
with various optimizers which they supplemented with
new spectroscopic and photometric data for $4U 1538-52$. This allows them to calculate an improved value
for the neutron star masses. Their derived values are $(0.87\pm0.07 M_{\odot})$ for 4U 1538-52
$(1.77\pm.08M_{\odot})$ for Vela X-1 and $(1.49\pm0.08M_{\odot})$ for Cen X-3. Freire et al.\cite{freire} obtained the mass of the compact star PSRJ 1614-2230 as $(1.97\pm0.04M_{\odot})$. Gangopadhyay et al.\cite{gan} estimated the radius of the compact star $7.866\pm0.21$ for 4U~1538-52, $9.69\pm0.2$ for
PSR~J1614-2230
$9.56\pm0.08$ for Vela~X-1 and
$9.178\pm0.13$ for Cen-X3. By using suitable values of the constants we have calculated the mass and radius of above mentioned compact stars which is presented in table $2$ and it is clear that calculated values of mass and radius of these stars is well fitted with the observational data. Moreover we have obtained the central density, surface density central pressure, compactness and surface redshift in table~3. The central density is in the order of $10^{15}g cm^{-3}$, surface density is in the order of $10^{14}~g cm^{-3}$, compactness factor is $<4/9$ and surface redshift $z_s\leq1.$\par
Hence we conclude that our proposed model may be used to
describe the interior of a superdense star corresponding to the exterior Schwarzschild line element.

\end{document}